\documentstyle[prl,multicol,aps,epsfig]{revtex}
\begin{document}
\title{Liquid, Glass  and Crystal in Two-dimensional Hard disks}
\author{Ludger Santen and Werner Krauth
\footnote{santen@lps.ens.fr; krauth@lps.ens.fr,
http://www.lps.ens.fr/$\tilde{\;}$krauth}
}
\address{CNRS-Laboratoire de Physique Statistique \\
Ecole Normale Sup{\'{e}}rieure,
24, rue Lhomond, 75231 Paris Cedex 05, France}
\maketitle
\begin{abstract}                
We study the thermodynamic and dynamic phase transitions in
two-dimensional polydisperse hard disks using Monte Carlo methods.
A conventional local Monte Carlo algorithm allows us to observe a
dynamic liquid-glass transition at a density $\rho^G$, which depends
very little on the degree of polydispersity. We furthermore apply Monte
Carlo methods which sample the Boltzmann equilibrium distribution at
any value of the density and polydispersity, and remain ergodic even
far within the glass. We find that the dynamical transition at $\rho^G$
is not accompanied by a thermodynamic transition in this two-dimensional
system so that the glass is thermodynamically identical to the  liquid.
Moreover, we scrutinize the polydispersity-driven transition from
the crystal into  the disordered phase (liquid or glass).  Our results
indicate the presence of a continuous (Kosterlitz-Thouless type)
transition upon increase of the polydispersity.

\end{abstract}

\begin{multicols}{2}
\narrowtext
For several decades now, enormous interest has been concentrated
on the glass transition in classical liquids \cite{review}. At this
transition, the viscosity of a structurally disordered liquid increases
dramatically and reaches values typical of a crystalline solid. The
underlying reasons for the spectacular slowdown of the system's time
evolution have been hotly debated and views very much  diverge on
this question \cite{Kauzmann,Gibbs,Mezard,Nagel}. The transition is
naturally related to the loss of thermal equilibrium during the quench
towards low temperatures, and it is well-known that the properties of a
glass strongly depend  on the detailed quench procedure \cite{Vollmayr}.
The theoretical description of this phenomenon in all its complexity is
a very difficult task.

Notwithstanding the intimate relationship of the glass transition with
\textit{loss} of thermal equilibrium ({\it i.e.} ergodicity breaking), it is
nevertheless possible to describe the glass \textit{within} thermal
equilibrium. This means that one studies configurations of the system
taken from the Boltzmann distribution (which exists in a
mathematical sense at any temperature and pressure) and computes their
thermodynamic and dynamic properties.  For any thermodynamically stable
glass, the reference to the quench history can thus be omitted,
and the theoretical analysis is much simpler.

The most serious problem with the `equilibrium glass' is of course that
the Boltzmann distribution is experimentally  inaccessible inside the
glassy phase. Attempts to characterize the true equilibrium properties of a glass necessarily rely on the extrapolation of experimental data.

However, computational methods have been developed which allow to bypass
this obstacle:  non-local Monte Carlo (MC) algorithms  \textit{can} remain
ergodic in the glassy phase for prominent model systems, such as
hard spheres, and allow to compute their equilibrium observables directly
\cite{nature}. With this  approach, we may address questions at the heart
of long-standing controversies, for example about the existence of an
inaccessible thermodynamic phase transition within the glass, which
has been rendered responsible for the dynamic transition 
\cite{Kauzmann,Gibbs,Mezard,Nagel}.

In the past, many classical statistical physics models have been
considered in the study of glasses. For a long time, it was believed
that the very simplest model systems (monodisperse hard spheres, hard
disks, and central potential models) did also possess a glassy phase
\cite{mono_glass}. However, more recent MC simulations have
shown that monodisperse hard spheres always crystallize, and that the
previously observed behavior was mainly due to finite-size effects
and limited computer resources \cite{Rintoul}.  This  has also been
confirmed in experiments with colloidal systems, where crystallization
preempts the glass transition for monodisperse hard spheres. In
order to avoid parasitic effects, some of the experiments have been
performed in  zero-gravity environments \cite{Russel}.

The situation changes abruptly, both in experiment \cite{Megen}
and in the simulations, in the presence of a small dispersity in
size. Several investigations have shown the existence of a so-called
`terminal polydispersity' of a few percent ({\it cf, e.g.,}
\cite{Kofke,Bartlett}), above which crystallization
can be avoided for any value of the external pressure, {\it i.e.}, up to
very high densities.

Polydispersity opens up opportunities to study two very interesting,
yet distinct phenomena: $(i)$ the glass transition, which is no longer
preempted by crystallization at sizeable values of the polydispersity;
$(ii)$ the thermodynamic transition of the crystal into a disordered state
(liquid or glass) upon an increase of the polydispersity.

In the present work, we study these two phenomena in two-dimensional
hard disks.  We investigate their dynamical and thermodynamical phase
diagram as a function of density $\rho$ (or pressure $P$) and of the
polydispersity for a given functional form of the size dispersion.
At polydispersities for which crystallization is impossible, we find a
purely dynamic liquid-glass transition  at a density $\rho^G$, which
depends very little on the degree of polydispersity $\varepsilon$.
For moderate values of $\varepsilon$, the glass transition density can
be defined even better than in the highly polydisperse case considered
earlier \cite{nature}.  Interestingly, $\rho^G$ is considerably higher
than the crystallization density for the monodisperse system.

Our vastly improved equilibrium simulation methods allow us to scan the
complete parameter space ($\varepsilon,\rho$).  Even far inside the glassy
phase, we can compute the equation of state and the compressibility
with very high precision.  The polydispersity-driven transition of the
crystal into the disordered phase (liquid or glass) also turns out to
be particularly interesting: we find strong evidence that this
disorder-driven phase transition is continuous.  Our findings can hardly
be reconciled with the presence of a first-order transition, and differ
from what was stated in a closely related work \cite{Sadr}.

In principle, for a polydisperse mixture of particles, one should choose the
particle sizes $r_i$ independently from a given probability distribution
$p(r)$. We would then have to perform an ensemble average over the
distribution of radii. In dynamical simulations, this average corresponds
to ensemble-averaging time correlation functions of a given sample. In
thermodynamic calculations, the average over radii can be incorporated
directly into the MC evaluation of the partition function,
as some authors have done \cite{Kofke}.

We have adapted a much simpler, yet practically equivalent approach, by
considering a `fixed probability increment' model for any probability
distribution $p(r)$ with $r>0$, where we pick the $N$ radii $r_i$
{\it e.g.} according to the following equation
\begin{equation}
\int_{0}^{r_i} dr  p(r) = \frac{i}{N+1}   
\label{fixed_probability}
\end{equation}
(with $\int_0^{\infty} dr p(r) = 1$).  In this way,
sample-to-sample fluctuations 
are completely eliminated, and the
most representative sample is generated for an arbitrary 
distribution $p(r)$. 

Here, we study  a model with a flat distribution of radii:
\begin{equation}
\label{flat_prob} 
  p(r) = \left\{ 
\begin{array}{ll}
      \mbox{const} &  r_{min} \leq r \leq r_{max}     \\
      0     &  \mbox{otherwise},
\end{array}          
\right.
\label{pdist}
\end{equation}
{\it i.e.} the radii are given by $r_i = r_1 + \Delta (i-1) $ (with $\Delta =
(r_n-r_1)/(N-1)$). As in \cite{Lee}, we set the total
particle volume as $V_{part} = N \pi/4$ and the Boltzmann factor as
$\beta=1$.

We define the polydispersity by the normalized width
of the distribution eq.(\ref{pdist})
\begin{equation}
\varepsilon  = \sqrt{<\delta r^2> /<r>^2}.
\label{ratio}
\end{equation}
Notice that the distribution eq.(\ref{pdist}) satisfies 
$\varepsilon  \le 1/\sqrt{3}$, this limit
being attained for $r_{min} = 0$.

Our simulations are performed in a periodic box of size $L \times 
\sqrt{3}L/2$, as is customary \cite{Lee}, and the only difference in
set-up between our thermodynamic and dynamic simulations is the choice
of ensembles. The equation of state $V(P)$ is most naturally computed in the
[NP] ensemble, where the size of the box is allowed to fluctuate.
The dynamical calculations are done in the [NV] ensemble, where the
time evolution of states can be best monitored. For these calculations,
a conventional local MC algorithm is used. Up to a trivial
rescaling of time, we expect this algorithm in the long-time limit to
give the same results as a molecular dynamics method.

The equilibration of polydisperse systems is a difficult task, because the
properties of the system depend sensitively on the relative arrangements
of particles, which have to be averaged over. Therefore, one needs an
algorithm that moves particles over large distances and averages over
different possible arrangements.  In the fluid phase this can in principle
be achieved  by a long series of local moves. In the crystal or the glass,
it is however mandatory to equilibrate the system by non-local moves,
as in the pivot cluster algorithm \cite{Dress}.

A polydisperse system with a continuous distribution of radii is easier
to simulate than a binary distribution, because it is often possible
to exchange particles with similar radii. These swaps also implement
non-trivial long-range moves, and have to be combined with a local Monte
Carlo algorithm.  The acceptance probability for such swaps decreases
only at the most extreme densities, when each particle is so close to
its neighbors that it can virtually never be replaced by a particle with
slightly larger radius.

In high-density {\em binary} mixtures ({\it cf} \cite{Grigera}), the pivot
cluster algorithm is clearly more appropriate, because the direct particle
exchanges freeze out.  In the present {\em polydisperse} case, however,
we have obtained identical results with both algorithms for systems of
$N=256$ and $N=1024$ particles even slightly below what we believe to
be close packing.

Extensive tests of ergodicity ({\it cf} \cite{nature}) fully confirm the
above statements: In very small systems ($N=15$), where the minimal
size differences between particles are larger, we can show that the
direct swap algorithm falls out of equilibrium at slightly {\em lower}
densities than the pivot cluster method for the above-mentioned reasons.

We point out a completely unrelated, yet crucial, issue: To compute
the equation of state at constant pressure, we need to include volume
changes into the MC algorithm. Extremely time-consuming sampling of the
volume by trial and error (using the Metropolis algorithm \cite{Lee})
can be avoided \cite{Kofke,prepare,footsample}.

\begin{figure}[htbp]
\centerline{
\epsfig{figure=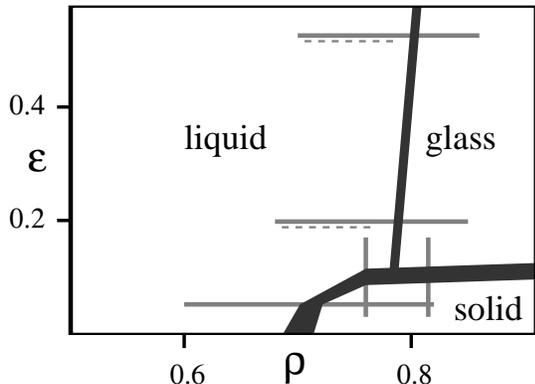,width=7cm} }
\caption{
Schematic phase diagram of the two-dimensional hard-disk system  
as function  of polydispersity $\varepsilon$ and density $\rho$. 
The dark lines indicate estimated phase boundary (obtained by 
linear interpolation of data points). The gray lines correspond
to the paths along which calculations were performed. 
Full lines: thermodynamic calculations (non-local MC algorithm);
broken lines: dynamic calculations (local MC).
}
\label{phasediagram}
\end{figure}

For orientation, we show in figure~\ref{phasediagram} the schematic
phase diagram as a function of density $\rho$ and polydispersity
$\varepsilon$ ({\it cf} eq.(\ref{ratio})) which results from our
thermodynamic and dynamic calculation, and which we discuss in the
remainder of this paper.  We have studied this diagram by the routes
indicated by gray lines in the figure, using either the non-local
MC algorithm (for thermodynamic averages), or the local 
MC method (to study the  dynamics).

\begin{figure}[htbp]
\centerline{
\epsfig{figure=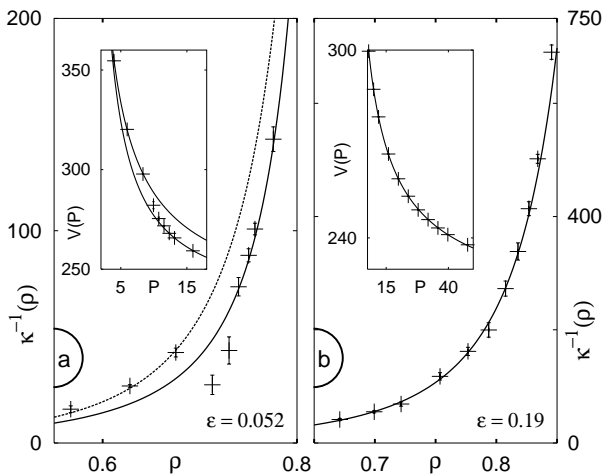,width=\columnwidth} }
\caption{
Inverse compressibility $\kappa^{-1}$ as a function of density $\rho$
and equation of  state $V(P)$ for two values of the polydispersity
$\varepsilon$ at $N=256$. Left: $\varepsilon = 0.052$: a thermodynamic
phase transition is clearly detectable, both in $V(P)$ and in
$\kappa^{-1}$.  Right:  $\varepsilon=0.19$: No thermodynamic phase
transition can be detected.  $\kappa$ is obtained by derivation of the
curve $V(P)$ {\em and} by direct computation from the fluctuations of
the volume.  }
\label{fig:eqs}
\end{figure}

In Figure~\ref{fig:eqs}a, we show the liquid and solid branches of the
thermodynamic equation of state as a function of pressure $P$ for a
fixed  $\varepsilon = 0.052$. The presence of a thermodynamic transition
is evident.

In these calculations at $\varepsilon=0.052$, and at $N=256$, we are
able to establish coexistence of the two phases as in the monodisperse
system \cite{Lee}. Even there however, the $N\rightarrow \infty$ limit
has remained controversial, and it is  not yet firmly established whether
the liquid-solid transition in 2-d monodisperse disks is first-order
or continuous (according to the KTNHY scenario, {\it cf} \cite{Strandburg}).
Notice that the transition shows up very clearly both in the 
equation of state and in
the compressibility, shown in the main figure~\ref{fig:eqs}a.

In figure~\ref{fig:eqs}b, we show analogous data for
$\varepsilon=0.19$. There, we see no indication of a thermodynamic phase
transition in the whole range of densities studied.  Similar results
were obtained at $\varepsilon=0.52$ (remember that $\varepsilon < 0.58$
{\it cf} text following eq.(\ref{ratio})); 
these data confirm and considerably extend our earlier
calculations \cite{nature}.  Within our flat probability distribution
of the radii of particles, there exists a  maximal polydispersity
$\varepsilon=0.58$, which corresponds to the limit $r_{min} \rightarrow 0$
in eq.(\ref{pdist}).

Our calculations thus confirm the existence of a terminal polydispersity,
which was studied mostly in three dimensions, but also evoked in related
two-dimensional work \cite{Sadr}.

\begin{figure}[htbp]
\centerline{\psfig{figure=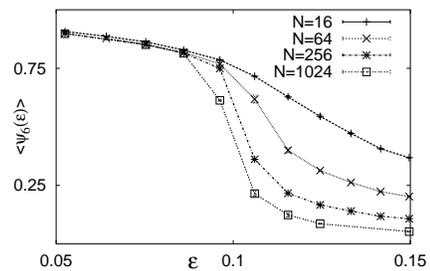,width=6cm}}
\caption{\protect{Average bond-orientational order-parameter for 
$P=17.1$. For polydispersities $\varepsilon \leq 0.08$,
the measurements converge towards a size-independent value, while we
expect $<\Psi_6> \to 0$ for larger polydispersities. The probability
distributions of $\Psi_6$ are unimodal, and give no indication of a
first-order phase transition.  Notice the smoothness of the finite-size
behavior, which is due to our choice of the fixed-probability increment
model eq.(\ref{fixed_probability}).}} 
\label{fig:KT} 
\end{figure}

The horizontal sweeps in the phase diagram of figure~\ref{phasediagram}
bracket the disordered-to-crystalline phase boundary, which  must be
essentially independent of $\varepsilon$ at large $\rho$.  We further
studied the thermodynamic system as a function of polydispersity
at constant pressure, along the vertical paths indicated in
figure~\ref{phasediagram}.  There,  our data strongly favor
a continuous transition: We did not find hysteresis in $V(P)$, 
and detected no abrupt change
of the volume distribution function in the transition region.
We also computed the average bond-orientational order-parameter $<\Psi_6>$
({\it cf,  e.g.}, \cite{Weber} for definitions and a discussion).  
There also,
the probability distribution of $\Psi_6$ has a single peak, and is
perfectly reproduced, without any indications of hysteresis. We conclude
that the transition is continuous.  In figure~\ref{fig:KT}
we show  $<\Psi_6>$ as a function of polydispersity $\varepsilon$ for
different system sizes. A detailed, more rigorous study of the probable
two-step melting process in this system is left for further work.
Let us note that the numerical  analysis of the  standard 
Kosterlitz-Thouless transition has proven to be very subtle even 
in the prototypical XY-model \cite{Olsson}.

\begin{figure}[htbp]
\centerline{\epsfig{figure=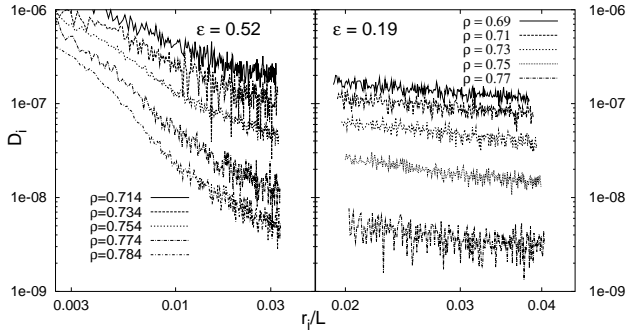,width=\columnwidth}}
\caption{Long-time diffusion constants as a function of particle size
$r_i/L$ for two values of the polydispersity  $\varepsilon$.  Left:
$\varepsilon = 0.52$, right: $\varepsilon = 0.19$. In both cases,
we witness a dramatic slow-down of the time-evolution as the density
$\rho^G \simeq 0.8$ is approached. This density  is much higher than the
liquid-solid transition density of the monodisperse hard-disk system,
but much smaller than the densities which we can reach (within thermal
equilibrium) with the non-local algorithms.
}
\label{fig:diff}
\end{figure}

Finally, we determine the dynamic properties of the system, in the
[NV] ensemble, with the local MC algorithm, and  at values of
$\varepsilon$ which ensure the absence of a liquid-solid transition.
Using the protocol of ref. \cite{nature}, we obtained the effective
diffusion constants $D_i$ ({\it cf} figure~\ref{fig:diff}), for 
$\varepsilon=0.19$
and $\varepsilon=0.52$. In both cases, the results are consistent with
the scaling form \cite{Fuchs}:
\begin{equation} 
D_i (\rho) \sim ( \rho^G_i -\rho)^{\alpha},
\end{equation} 
where we extrapolate the diffusion constants for each particle
separately.  The diffusion constants $D_i$ strongly depend on $i$, but
the extrapolated transition densities $\rho^G_i$ do not, especially for
small $\varepsilon$. We take these extrapolated values as a definition
of the glass transition density $\rho^G$. Only at  large $\varepsilon$
(left part of figure) can some very small particles ($r_{min}/r_{max}
= 19$) escape through slits left open in the system. This leads to
some additional structure in the $D_i$ plot on the left half of figure
\ref{fig:diff}, and to an increased value of the extrapolated $\rho_i^G$
for small $i$.

In conclusion, our study of the polydisperse hard disk system has
given the homogeneous phases shown in the schematic diagram figure
\ref{phasediagram}. Inhomogeneous phases \cite{Bartlett}  seem to play no
role at the parameters studied in this paper. A tendency towards phase
separation  should first show up in irregular finite-size behavior,
which we have not encountered.  We find it particularly significant
that the glass transition depends so little on polydispersity, and takes
place at densities (and pressures) much higher than the crystallization
density in the monodisperse system. Using our specialized MC methods,
we are able to probe the thermodynamics of the hard-disk system far
within the glassy phase, where we find no indications of an accompanying
thermodynamic transition.

Acknowledgements: We thank P. Le Doussal for helpful
discussions. L.~S. acknowledges support from the Deutsche 
Forschungsgemeinschaft under Grant No. SA864/1-2.

\end{multicols} 

\end{document}